\documentclass{article}
\usepackage{amsmath}
\usepackage{color}
\usepackage{amsfonts}
\usepackage{graphics}
\def\a{a}
\def\b{b}
\renewcommand\Re{{\rm Re}}
\def\W{\mathcal W}
\def\M{\mathcal M}

\def\e{\eta}
\def\ex{e^\xi}

\def\x{{\bf x}}
\def\y{{\bf y}}
\def\k{{\bf k}}

\def\bea{\begin{eqnarray}}
\def\eea{\end{eqnarray}}

\def\rdq{\textquotedblright}

\begin{document}

\title{ \bf A note on canonical quantization of fields on a manifold}
\author{Ugo Moschella$^{1,2}$ and Richard Schaeffer$^3$ \\$^1$Universit\`a dell'Insubria, 22100 Como, Italia, \\$^2$INFN, Sez. di Milano,
Italia\\$^3$Institut de Physique Th\'eorique, CEA - Saclay,
France}

\maketitle
\begin{abstract}
We propose a general construction of quantum states for linear canonical quantum fields on
a manifold, which encompasses and generalizes the ``standard'' procedures
existing in textbooks. Our method provides pure and mixed states on the same footing.
A large class of examples finds a simple and unified treatment in our approach.
Applications discussed here include thermodynamical equilibrium  states for Minkowski fields and
quantum field theory in the Rindler's  and in the open de Sitter universes.
Our  approach puts the above examples into perspective and unravels new possibilities for quantization.
We call our generalization ``extended canonical quantization''  as it is suited to attack  cases
not directly covered by the standard canonical approach.

\end{abstract}

\section{Introduction}

Switching from classical to quantum mechanics  relies on a set
of recipes that, despite their degree of arbitrariness, have
proven to successfully describe a large variety of physical phenomena.
In the simple case of a mechanical system having a finite number of degrees of
freedom,  the time-honored procedure of ``canonical quantization''
essentially amounts to replacing Poisson brackets by commutation relations:
\begin{eqnarray}
\{q_i.p_j\} = \delta_{ij} \longrightarrow [Q_i.P_j] = i \hbar \delta_{ij}, \ \ i,j = 1,\ldots, N. \label{cct}
\end{eqnarray}
In this step  ${Q_i}$ and ${P_j}$ are understood as elements of an abstract Heisenberg algebra.
Under suitable technical assumptions  the fundamental uniqueness theorem by Stone and von Neumann establishes that
there exists only one  representation of the commutation relations (\ref{cct}) by operators in a Hilbert space $\cal H$
\begin{equation}
Q_i \to \hat Q_i, \ \ P_j \to \hat P_j: \ \ [\hat Q_i,\hat P_j] = i \hbar \delta_{ij} {\bf 1}_{\cal H},
\end{equation}
all other  representations beeing unitarily equivalent; ${\bf 1}_{\cal H}$ denotes the identity operator in the Hilbert space $\cal H$.

The situation drastically changes when considering systems with infinitely many degrees of freedom.
The Stone-Von Neumann  theorem  fails for infinite systems and there exist
uncountably many inequivalent  Hilbert space  representations of the canonical
commutation relations \cite{strocchi,haag}. Therefore, quantizing an infinite system such as
a field  involves two\footnote{Also
in the case of infinite systems, one might want to reserve the name ``quantization''
only to the first step that consists in replacing a classical Poisson algebra
with a CCR or a local observable algebra. This is the viewpoint advocated in the algebraic approach to quantum field theory
and there are good reasons to subscribe to this (representation independent) notion of quantization.
Indeed, it may be that the algebra which one tries to construct has non-trivial ideals which are missed by proceeding directly to a representation \cite{Buchholz}.
We adopt however the usage that is (often tacitly) adopted
in the vast majority of quantum field theory textbooks
to call ``quantization'' a Hilbert space realization of the field algebra.
Indeed,  the standard  machinery of
quantum field theory can be put at work only in a concrete Hilbert space realization (i.e. computing Feynman propagators,
constructing perturbative amplitudes,  etc.)} distinct steps:
\begin{enumerate}
\item construction of an infinite dimensional algebra describing the degrees of freedom of the quantum system;
\item construction of a Hilbert space representation of that algebra.
\end{enumerate}
Unfortunately, a complete classification of the possible representations of
the canonical commutation relations does not  exist and is not foreseen in the near future.
This lack of knowledge is especially {relevant} in curved
backgrounds where, generally speaking, the selection of a fundamental state {cannot be
 guided by the same physical principles as in flat space}.  Indeed, while the
CCR's have a purely kinematical content, the construction
and/or the choice of one specific representation {in a Hilbert space}
is always
related to dynamics and different dynamical behaviors require
inequivalent representations of the CCR's (see e.g.
\cite{strocchi}). This is related to many fundamental issues
such as renormalizability, thermodynamical equilibrium and
entropy, phase transitions, etc..

The above considerations suggest that
it might be useful to re-examine
once more the {quantum theory} of a free Klein-Gordon
(KG) field on a curved  background. Another reason for doing so is the occurrence \cite{sasaki,MS} of situations where one is forced to
stretch the canonical formalism beyond its limits to get a
physically meaningful result.

In flat spacetime the ``standard'' quantization
{(i.e. Hilbert space realization)} of a field is
founded on the  physical requirement to
retain only the positive frequency solutions of the Klein-Gordon equation in
order to construct the N-particle excitations of the vacuum.
This choice is called the ``spectral condition''; in the
general setup of QFT it is an independent assumption
\cite{SWight} and produces a construction that is so natural
that the non-uniqueness of the ensuing  quantization is hardly
mentioned in standard textbooks of QFT and often overlooked and
forgotten.

The situation is {vastly different for} fields on curved
spacetimes; here a global energy operator may not exist at all
and a true spectral condition is in general not available.
What remains is a well-developed
formalism for canonical quantization of
linear fields\footnote{See e.g. the classic reference book \cite{Birrell};
a more mathematical description of the method can be found in \cite{wald} and a recent short account in \cite{kay3}.},
based on
the introduction of a conserved inner product
{in the space of classical solutions of the Klein-Gordon
equation: given two complex solutions $\phi_1$ and $\phi_2$ their inner product is defined as the integral of the conserved
current
$
j_\mu(x) = i {\phi^*_1(x)} \mathop{{\nabla}_\mu} \phi_2(x) -
 i \phi_2(x) \mathop{{\nabla}_\mu}{\phi^*_1(x)}$
 over a three-dimensional Cauchy surface; the necessary assumption
is therefore the global hyperbolicity of the spacetime manifold (see e.g. \cite{wald}).
A canonical quantization is then
achieved by finding a splitting of the above space of complex solutions
into the direct sum of a subspace where the inner product is positive
and of its complex conjugate.
The ``1-particle'' Hilbert space of the theory is finally obtained
by completing the chosen positive subspace in the Hilbert topology defined by the inner product.
Of course, infinitely many inequivalent theories can be constructed this way:
for example, once given a certain quantization,
a whole family of inequivalent alternatives can be obtained
by means of Bogoliubov's transformations.

A considerable amount of work has been devoted to the attempt
of formulating various alternative  prescriptions
to select, among the possible representations of a field theory, those which
can have a meaningful physical interpretation:
here we quote,
in view of their importance in semiclassical general relativity, only
{\em the local Hadamard condition}, (\cite{wald,kay3,kayw} and references therein)
and the {\em microlocal spectral condition} \cite{micro}.

However, before facing the problem of {selecting  a quantization by its physical properties, one should better
 make sure not to have omitted relevant possibilities;
to this scope  it is important to construct the most general available quantization scheme.
The contribution of the present note is specifically at this level} and
starts from the rather obvious remark
that Bogoliubov transformations are not enough to produce the most general
class of canonical theories constructible starting from a given
quantization. This observation opens the door to a useful
generalization of the canonical formalism.  The latter provides access to
many additional inequivalent quantizations,
some of them having} a potentially important physical
significance.

{One important point in our construction is that it allows for a
more flexible use of coordinate systems.
The standard formalism of canonical quantization, that we have already briefly summarized,
requires the global hyperbolicity of the
spacetime manifold. A globally hyperbolic  manifold
can be foliated by a family of Cauchy surfaces
$\Sigma_t$ where $t$ is a temporal coordinate; the Klein-Gordon inner product
is built by integrating the current associated with two complex solutions over any Cauchy surface
and its value does not depend on the chosen surface.

{However, it is not uncommon in concrete examples to
make use of coordinates  that cover only a
portion of a certain spacetime manifold and that the
spacelike  hypersurfaces defined by a condition of the form $t=${\em constant} in the given coordinate system
are not Cauchy surfaces for the extended manifold;
the patch covered by the coordinate system may or may not
be a globally hyperbolic manifold in itself.

For example, this situation is encountered
in black-hole spacetimes or in the Rindler coordinate system
of a wedge of a Minkowski spacetime, and this is well-known and understood \cite{hawking,unruh,sewell,bisognano,kay,kay2,fulling,kayw}.
The same phenomenon may also happen in cosmological backgrounds.
Depending on the behavior of the scale factor $a(t)$, the surface  $t=${\em constant} may fail to
be a Cauchy surface for the maximally extended manifold
(while it is a Cauchy surface for the patch covered by the Friedmann-Lema\^{\i}tre-Robertson-Walker coordinates).
In this circumstances textbooks suggest  the
use of the standard canonical formalism \cite{Birrell}, but many possible
quantizations are lost in this very initial step.
We  will see how our generalization of the canonical formalism will
allow to construct representations that are not attainable
}{by the standard recipes in the above circumstances.}

Our construction is useful already in  flat spacetime; in
Section \ref{sect:baba} we show how the KMS quantization of the
free scalar field fits in our generalization of the canonical
formalism. In
 Sect. \ref{sect:Rin&Min} we will give a fresh
look to the Rindler model and the Unruh effect and in
Section \ref{sect:opendeSitt} to the open de Sitter model. The
common feature shared by our discussion of these well-known
examples is that the Minkowski Wightman vacuum and the
preferred Euclidean de Sitter vacuum are here reconstructed
working solely inside the patches covered by the relevant
coordinate systems.

{As a conclusion, we summarize our findings in Sect.
\ref{sec:concl}.}

\section{A general canonical scalar quantum  field \label{sect:phidetout}}

Let us consider a real scalar quantum field $\phi(x)$ on a
general background ${\cal M}$.
At this initial level of generality it is not necessary to assume any equation of motion for the field $\phi$.
From a mathematical viewpoint, the field is a map
\begin{equation}
 f \to \phi(f) = \int \phi(x) f(x) dx
\end{equation}
from a suitable linear space of test functions, say ${\cal D}(\cal M$), to a corresponding field
algebra $\cal F$.
The commutation relations have a purely algebraic
content; in particular, {\em for generalized free fields} the commutator is a
c-number, i.e. a multiple of the identity element of $\cal F$:
\begin{equation}[\phi(f), \phi(g)] = C(f,g)\, {\mathbf 1} = \int C(x,x')f(x) g(x') \, dx \, dx'\,  {\mathbf 1};
\label{comm}\end{equation}
in this equation $C(x,x')$ is an {antisymmetric} bidistribution on the manifold ${\cal M}$ which
has to vanish coherently with the notion of locality inherent
to ${\cal M}$, i.e.
\begin{equation}
C(x,x') = 0 \ \ \ \makebox{ for }  x,x' \in {\M} \ \ \makebox{``spacelike separated'';}
\end{equation} $dx$ shortly denotes  the invariant volume form. {The covariant formulation used here supersedes
the ``equal time'' CCR's
mentioned  in the Introduction\footnote{See \cite{SWight} for a discussion on this point.
For  Klein-Gordon fields  the covariant commutation relations
and the equal time CCR's are equivalent.}.}

{The next step is to realize
the field as an operator-valued distribution in a
Hilbert space $\cal H$,
\begin{equation}
\phi (f) \longrightarrow \widehat \phi (f),
\end{equation}
and we know  that there are infinitely many inequivalent
such representations,
having different physical interpretations or
no interpretation at all.}

In the following, we will restrict our attention to generalized free fields,
i.e. fields whose truncated $n$-point functions vanish for $n>2$ (and the one-point function vanishes as well). The quantum theory
of   generalized free fields is therefore completed encoded in
the knowledge of a positive
semi-definite\footnote{This hypothesis should however be relaxed to deal with local and covariant gauge quantum field theories \cite{strocchi}} two-point function ${\W}(x,x')$, a distribution
whose interpretation is that of being the two-point ``vacuum'' expectation value of the field:
\begin{equation}
{\W}(x,x') \equiv \langle \Omega, \widehat \phi(x) \widehat \phi (x') \Omega\rangle.
\end{equation}
The following is the crucial property that ${\W}(x,x')$ has to satisfy to induce a representation of the commutation rules (\ref{comm}):
${\W}(x,x')$ must  realize a splitting of the commutator $C(x,x')$ by solving the following fundamental functional equation:
\begin{equation}
{C}(x,x') = {\W}(x,x')- {\W}(x',x).  \label{CR}
\end{equation}
Given a $\W$ satisfying (\ref{CR}), a Hilbert space representation
of the field algebra (\ref{comm}) can be
constructed explicitly.
The one-particle space ${\cal H}^{(1)}$ is obtained by the standard Hilbert space completion
the space of test function ${\cal D}(\cal M$) w.r.t.
the positive semi-definite pre-Hilbert product provided by the two-point function (see e.g. \cite{SWight}):
\begin{equation}
\langle f, g\rangle =
\W(f^*,g) = \int {\W}(x,x')f^*(x) g(x') \, dx \, dx'.
\end{equation}
The full Hilbert space of the theory is the  symmetric Fock
space ${\cal H}=  F_s({\cal H}^{(1)})$.
Each field operator $\widehat\phi(f)$ can be decomposed into
``creation'' and ``annihilation'' operators
$\widehat\phi(f)=\widehat\phi^{+}(f)+\widehat\phi^{-}(f)$ defined by their action on the dense subset
of  $F_s({\cal H}^{(1)})$ of elements the form ${h} = ({h}_0, {h}_1,\ldots
{h}_n,\ldots,0, 0,0,\ldots)$:
\begin{equation}
\left({\widehat{\phi}}^{-}(f){h}\right)_{n}(x_1,\ldots,x_n)=
{\sqrt{n+1}}\int {\W}(x,x') f(x)
{h}_{n+1}(x',x_1,\ldots,x_n)
dx dx' ,
\end{equation}
\begin{equation}
\left(\widehat\phi^{+}(f){h}\right)_n(x_1,\ldots,x_n)=\frac{1}{\sqrt{n}}
\sum_{k=1}^{n}f(x_k){h}_{n-1}(x_1,\ldots, \hat{x}_{k},\ldots,x_n).
\label{azzo}
\end{equation}
Because of (\ref{CR}) these formulae imply the commutation relations (\ref{comm}).
\subsection{Klein-Gordon fields}
We now proceed to describe the construction of
two-point functions $\W$ that fulfill the above requirements
by further restricting our attention to Klein-Gordon fields.

The most general two-point function
will be shortly seen to be obtainable beyond
the scheme of canonical quantization, but let us
follow at first the standard recipes \cite{Birrell} and construct a family of complex
classical solutions $\{u_i(x)\}$  of the Klein-Gordon equation;  $\{u_i(x)\}$  has to be a complete
(see e.g. \cite{fulling}) and orthonormal set
in the following sense\footnote{There are mathematical problems in this very initial step because
the space of complex solutions of the Klein-Gordon equation is not an Hilbert space but it has just an indefinite metric induced by the Klein-Gordon product; therefore one is not entitled to speak of a ``basis'' unless some Hilbert topology is added (but  extra information is needed). This problem has been circumvented and solved \cite{wald,kayw} by studying the possible Hilbert topologies that one can give to the space of {\em real} classical solutions of the Klein-Gordon equation for a description of these results). Here we are interested in the  heuristics, but there is a relation of our findings with the aforementioned  construction. This will be studied elsewhere.}
\begin{equation}
(u_i,u_j) = \delta_{ij},\;\;\;\;( u^*_i,  u^*_j) = -\delta_{ij},
\;\;\;\;(  u_i, u^*_j) = 0, \label{kgbase}
\end{equation}
where $(u,v)$ denotes the Klein-Gordon inner product on a globally hyperbolic manifold $\cal M$.

The  standard  canonical quantization of the Klein-Gordon field corresponding to the set $\{u_i(x)\}$
is then achieved in the following two steps: first step, write the formal expansion of the field
\begin{equation}
\phi(x) = \sum[ u_i(x)a_i + u^*_i(x)a_i^\dagger]
\end{equation}
in terms of the elements of a CCR algebra
\begin{eqnarray}
&& [a_i,a^\dagger_j] = \delta_{ij}, \ \ \ \ \ \ [a_i,a_j]=0 , \ \ \ \ \ \ [a^\dagger_i,a^\dagger_j]= 0;
\end{eqnarray}
second step, construct the corresponding Fock representation $\phi(x) \to \widehat \phi(x)$
which is fully characterized by the annihilation conditions
\begin{equation}
\widehat a_i |\Omega\rangle = 0,     \ \ \ \forall i.
\end{equation}
While the first step just encodes the covariant (unequal time) commutation relations
\begin{equation}
[\phi(x),\phi(y)] = C(x,y) = \sum [u_i(x) u_i^*(y) - u_i(y) u_i^*(x)]; \label{commu}
\end{equation}
 the Fock space construction in the second step is completely equivalent to the assignment
of the two point vacuum expectation value
\begin{equation}
\W(x,x') = \langle \Omega, \widehat \phi(x) \widehat \phi(y)  \Omega \rangle = \sum u_i(x) u_i^*(y). \label{standard}
\end{equation}
Note that this two-point function is the most simple solution
for the  split equation (\ref{CR}), after the covariant commutator $C(x,y)$
has been expanded in the basis of modes ${u_i(x)}$ as in Eq. (\ref{commu}); the permuted function is simply
\begin{equation}
\W'(x,x') = \W(x',x) = \langle \psi_0, \widehat \phi(x') \widehat \phi(x)  \psi_0 \rangle = \sum  u_i^*(x)u_i(x').
\end{equation}

There are however infinitely many other solutions of the functional equation (\ref{CR}) giving rise to (possibly) inequivalent
canonical quantizations. We will now show how to construct many of them
by using the complete set (\ref{kgbase}); to this end, it is useful to
begin the discussion  by reviewing the standard theory of Bogoliubov transformations.

A Bogoliubov transformation amounts to the construction of a second complete system $\{v_i(x)\}$  by
the specification of two complex operators (matrices) $\a_{ij}$ and $\b_{ij}$ such that
\begin{eqnarray}
v_i(x) &=& a_{ij} u_j(x) + b_{ij}  {u}^*_j(x) ,\label{bogdir}\\ u_j(x) &=&  v_i(x)\, a_{ij}^*  - {v}^*_i(x)\, b_{ij} .
\end{eqnarray}
By composing  the direct and inverse transformations it follows  that
 $a$ and $b$ must satisfy the following conditions:
\begin{equation}
a_{il} a^*_{jl} - b_{il} b^*_{jl} = \delta_{ij}, \ \ \ \ \ \ \ \  a_{il} b_{jl} - b_{il} a_{jl} = 0,
\end{equation}
\begin{equation}
a_{li}^* a_{lj}  -b_{li}  b^*_{lj} = \delta_{ij}, \ \ \ \ \ \ \   a^*_{li} b_{lj} - b_{li}
a^*_{lj} = 0. \label{bb}
\end{equation}
The standard Fock quantization based on the system $\{v_i(x)\}$ is then encoded in the two-point
function
\begin{eqnarray}
\W_{\a,\b}(x,x') = \sum v_i(x) v_i^*(y) =  \sum\, [\a_{ij
} \, \a^*_{il} u_j(x)  u^*_l(x') + \b_{ij} \b^*_{il} \, u^*_j(x) \, u_l(x') \cr + \a_{ij
} \b^*_{il}\,  u_j(x)  u_l(x') + \b^*_{ij} \a_{il} \, u^*_j(x) \, u^*_l(x')]   \label{kgab}
\end{eqnarray}
interpreted as the two-point vacuum expectation value of the quantum field.
Positive definiteness of (\ref{kgab}) is evident.

When  $b$ is a Hilbert-Schmidt operator this quantization turns out to be unitarily equivalent
to the Fock quantization (\ref{standard}). Otherwise (\ref{standard}) and (\ref{kgab}) give rise to inequivalent quantizations.
The commutator must however be independent of the choice of
$a$ and $b$; condition (\ref{bb}) precisely implies that this is true:
\begin{eqnarray}
\W_{\a,\b}(x,x')- \W_{\a,\b}(x',x) = \sum [v_i(x) v^*_i(x')-v_i(x')
v^*_i(x)] = \cr =\sum [u_i(x) u^*_i(x')-u_i(x')
u^*_i(x)] = C(x,x'). \label{cancomm} \end{eqnarray}

At this point in most
textbooks the story about canonical quantization comes to an
end. {There is however is  much room left}. We show in this paper that new representations
can be produced by enlarging the family of two-point functions displayed in (\ref{kgab}).
Consider indeed two hermitian matrices $A$ and $B$ and a
complex matrix $C$ and construct  the general quadratic form
\begin{eqnarray}
Q(x,x') =  \sum [A_{ij} \, u_i(x) u^*_j(x') + B_{ij} \, u^*_i(x)
u_j(x')+  \cr + C_{ij} u_i(x) u_j(x') + C^*_{ij} u^*_i(x)
u^*_j(x')].\end{eqnarray}
Now we ask that $Q(x,y)$ be a solution of Eq. (\ref{CR}); by imposing the commutation relations (\ref{commu})
we get the following conditions on the operators $A, B$ and $C$
\begin{equation}
A_{ij} -
B_{ji} = \delta_{ij}, \ \ \ \ C_{ij} - C_{ji} = 0.
\end{equation}
In the end we obtain the most general expression for a canonical two-point function
solving the Klein-Gordon equation:
\begin{eqnarray}
\W(x,y) &=& \sum  \left[ \delta_{ij} + B_{ji}\right] u_i(x) u^*_j(x') + \sum B_{ij} u^*_i(x) u_j(x') \cr
&+&  \Re \sum C_{ij}
[u_i(x) u_j(x') + u_i(x') u_j(x)] + S(x,x'). \label{general}
\end{eqnarray}
Only the first diagonal term at the RHS
contributes to the commutator. The other terms
altogether constitute the most general combination of the modes
(\ref{kgbase}) so that the
total contribution to the commutator vanish.
Eq. (\ref{general}) provides a considerable enlargement of the family
of possible quantizations as compared to the subset (\ref{kgab}) provided by the
standard canonical quantization rules plus Bogoliubov transformations.
We stress once more that however ``canonicity''
is preserved in the sense the commutator is always the same and does not depend on the operators $B$ and $C$.
{For example, when $\cal M$ is Minkowski space,
Eq. (\ref{general}) is the most general
superposition of  positive  and  negative energy modes
that preserves  the standard equal-time CCR's.

Eq.(\ref{general}) reduces to a Bogoliubov
transformation of the reference theory only in the special case
(\ref{kgab}). These states are {\em pure states}.
The states that we have added in the enlarged canonical formalism
are in general mixed states:  the representation of the field algebra is not irreducible.

As it will made clear in the discussion of concrete examples, simple but important examples of mixed states
are provided by the following family of models:
\begin{equation}
\underline \W_{\a,\b}(x,x') = \sum [\a_{ij
}\a^*_{il} \, u_j(x) u^*_l(x') + \b_{ij}\b^*_{il} \, u^*_j(x) u_l(x')], \label{kgab2}
\end{equation}
with  $\a_{ij}\a^*_{il} - \b^*_{ij}\b_{il} = \delta_{jl}$.

{\em There is even place for a further generalization}: the  term
$S(x,x')$, which  we have not yet commented. This is a bisolution  of the Klein-Gordon equation
that is not ``square-integrable'' (even in a generalized sense).
It is of classical nature
and symmetric in the exchange of $x$ and $x'$.
Quantum constraints do not generally forbid the existence of
such a contribution. Its introduction  may be necessary to
access to degrees of freedom which cannot be described in terms
of the $L^2$ modes (\ref{kgbase}). This important extension { to non-$L^2$  ``classical''   modes}
deserves a thorough examination \cite{usexp} is
incidentally mentioned here.

We now pass to discussing a few examples where the virtues
of our extension of the canonical formalism will be made clear.

\section{KMS quantization \label{sect:baba}}

A
toy example in this class can be constructed  in flat spacetime
starting from the standard plane wave solution to
(\ref{kgbase}):
\begin{equation}
u_\k(x)=u_\k(t,\x) = \frac{1}{\sqrt{2\omega(2\pi)^3}} \exp(-i\omega t +i\k\cdot \x),\;\;\;\;\omega = \sqrt{\k^2+m^2}. \label{minkmodes}
\end{equation}
The reference two-point function  w.r.t this set of modes
satisfies positivity of the spectrum of the energy operator in
every Lorentz frame \cite{SWight}:
\begin{equation}
\W(x,x') = \int {u}_\k(x){u}^*_\k(x')d\k =
\frac{1}{(2\pi)^3}
\int e^{-ik(x-x')}
\theta(k^0)\delta(k^2-m^2)\,dk.
\end{equation}
The field $\phi$ and its conjugate   $\pi(x) =
\partial_t \phi(x)$ can be reconstructed as in (\ref{azzo}); CCR's hold literally i.e.
$[\phi(t,\x),\pi(t,\y)]  = i \delta({\x}-{\y}).$ In the toy
example that follows we consider the diagonal operators
\begin{equation}
\a_{\k\k'}  = \sqrt{\frac{e^{\beta/2}}{{2\sinh(\beta/2)}}}
\delta_{\k\k'},\;\;\;\;\b_{\k\k'} = \sqrt{\frac{e^{-(\beta/2)}}{{2\sinh(\beta/2)}}}\delta_{\k\k'},
\label{eleven}
\end{equation}
depending on a constant $\beta$. Proceeding as in (\ref{kgab})
by Bogoliubov transformations one gets noncovariant canonical
quantizations of the Klein-Gordon field. On the other hand
$\underline\W_{\a,\b}$ constructed as in (\ref{kgab2}) provides
a covariant canonical quantization that cannot be obtained by
Bogoliubov transformations. { Consequently the two-point
function $\underline\W_{\a,\b}$ cannot satisfy the spectral
condition \cite{SWight} and, as anticipated, states with negative energy are
now present in the Hilbert space of the model:}
\begin{equation}
\widetilde {\underline \W}_{\a,\b}(k) = \left[ \frac{1}{1-e^{-\beta }}\,
\theta(k^0) + \frac{1} {e^{\beta }-1}\,\theta(-k^0)\right]\delta(k^2-m^2). \label{wight}
\end{equation}

In the next example the { above toy model is generalized to non-constant}
but yet diagonal matrices. Let us consider in particular the
operators
\begin{equation}\a_{\k\k'} = \sqrt{\frac{e^{\beta \omega/2} }{{{2\sinh(\beta
\omega/2)}}}} \  \delta_{\k\k'}
 ,\;\;\;\b_{\k\k'} = \sqrt{\frac{e^{-\beta
\omega}}{{{2\sinh(\beta \omega/2)}}}} \ \delta_{\k\k'}.
\end{equation}
Here Bogoliubov transformations (\ref{kgab}) give an otherwise
uninteresting canonical Klein-Gordon quantum field theory. On the
contrary, the two-point function
\begin{equation}
{\underline \W}_{\beta}(x,x') =
\frac{1}{(2\pi)^3}
\int e^{-ik(x-y)}\left[ \frac{\theta(k^0)}{1-e^{-\beta k^0}}\, + \frac{\,\theta(-k^0)} {e^{-\beta k^0}-1}\right]\delta(k^2-m^2)dk   ,  \label{weight}
\end{equation}
{that is a special instance of the family of models exhibited in Eq. (\ref{kgab2}),}
is of fundamental importance in quantum field theory as it
provides the {\em well-known} Kubo-Martin-Schwinger (KMS) thermal representation
of the Klein-Gordon field at inverse temperature $\beta$
\cite{Birrell,kms,Bros:1992ey}.
{This quantization
can of course can be obtained by a variety of other means}.
To compare our construction to the literature, we see a point of contact
with the so-called thermofield theory \cite{umezawa,kay2} where the KMS
representation is also obtained in an approach inspired from
canonical quantization in a fundamental state rather than by a
statistical average as usual \cite{kms}. However, there is a
difference in that we do not need to introduce any doubling of
the degrees of freedoms { by means of an auxiliary ``dummy'' space}
but we insist in representing one and
the same field algebra.
Indeed the so called doubling of the degrees of freedoms {used in
thermofield  theory} is an artifact of the momentum space representation used  to implement the x-space CCR's; {these momentum space deformations however are a very useful and convenient mathematical tool to perform practical calculations.}
The KMS construction {thus is seen to be an example encompassed by the construction (\ref{general}), which of course is much more general.}

\section{Rindler space \label{sect:Rin&Min}}

In the following important example we will apply our method to
revisit the widely studied  Rindler spacetime and the Unruh effect
\cite{unruh}. To keep the discussion at the simplest level, but still rigorous
and general, we will consider the two-dimensional massive Klein-Gordon fleld.
Indeed, the  massless
case, which is usually discussed in textbooks (see e.g. \cite{Birrell}), is very special
because of its conformal invariance.
Also, the massless Klein-Gordon theory in two-dimensions
has an infrared behavior that renders the (local and covariant) canonical two-point function
not positive-definite and the
linear space of states of the model includes necessarily  negative-norm unphysical states \cite{klaiber,Morchio}.
The general dimensional case easily follows from the two-dimensional massive case.

The two-dimensional Rindler spacetime can be identified with (say)
the right wedge  of the two-dimensional Minkowski spacetime
w.r.t. a chosen origin. The relevant coordinate system is
constructed from the action of the Lorentz boosts that leave
the Rindler wedge invariant (which are of course isometries  of the wedge):
\begin{eqnarray}
&& x^0 = \ex \sinh \e,\;\;x^1 = \ex
\cosh \e, \;\;  \label{rinco} \\  && ds^2= e^{2 \xi }({d\eta}^2-d\xi^2).
\end{eqnarray}
The variable  $\eta$ is interpreted as the Rindler time coordinate.
With the help of these coordinates the massive Klein-Gordon equation is
written as follows:
\begin{equation}
\partial^2_\eta \phi-\partial^2_\xi \phi + m^2 e^{2\xi}\phi = 0.
\end{equation}
Let us consider factorized solutions of the form $ u(\eta,\xi) = e^{-i\omega
\eta}F_{\omega}(\xi) $, which are of positive frequency $\omega>0$ w.r.t. the Rindler time
$\eta$; the factor $F_{\omega}(\xi)$
obeys the modified Bessel equation:
\begin{equation}
-\partial^2_\xi F + m^2 e^{2\xi}F= \omega^2 F .
\end{equation}
 The solution that behaves
well at infinity is the Bessel-Macdonald function
 $K_{i\omega}(me^\xi)$ \cite{BAT2}; therefore, a convenient system solving (\ref{kgbase}) for the massive Klein-Gordon
equation in the Rindler universe can be written  as follows:
\begin{eqnarray}
\left\{\begin{array}{lll}u_\omega(\eta,\xi) &=&\frac{\sqrt{\sinh \pi \omega}}{\pi}
e^{-i\omega \eta}K_{i\omega}(me^\xi)\;\;\cr
u^*_\omega(\eta,\xi) &=&\frac{\sqrt{\sinh \pi \omega}}{\pi}
e^{+i\omega \eta}K_{i\omega}(me^\xi)\;\; \ \
\end{array} \right. \omega>0.\label{rinmodes}
\end{eqnarray}
A spacelike surface that may be used to compute the normalization in (\ref{rinmodes}) is for instance the
half-line $\eta= \eta_0$ ($\xi \in {\Bbb R}$).  The result does not depend on the choice of one
particular half line because they all share the same origin.
In doing this we
are applying the standard canonical formalism in the Rindler
wedge.

Of course the
system (\ref{rinmodes}) is not enough to perform canonical
quantization on the whole Minkowski spacetime.
In the original approach \cite{unruh,Birrell} the system (\ref{rinmodes}) is
supplemented by an analogous family of ``left'' modes.
The so completed system can be used to put
the machinery of Bogoliubov transformations at work and  recover the standard
Wightman ground state  \cite{unruh,Birrell}.
In the end, the Unruh effect is exhibited  by restriction of Wightman vacuum to the Rindler wedge (in the coordinate system (\ref{rinco})).  This fact is however general and model independent: restricting a Wightman quantum field theory to a wedge always gives rise to a KMS state \cite{sewell,bisognano}.

We are now going to show how our formalism allows for a direct
construction of the Wightman vacuum solely within the right Rindler wedge in terms of the
``right'' modes (\ref{rinmodes}) alone, avoiding the need of extending the system to the left wedge.

In the first step, insertion of the modes (\ref{rinmodes}) in Eq. (\ref{general}) (with $S(x,y)=0$)  provides a huge family of mathematically
admissible two-point functions (and therefore states) for the massive Rindler Klein-Gordon field,
all of them sharing the same commutator $C(x,y)$ and, a fortiori, the canonical equal time commutation relations.

In the second step, we select those theories
in which the wedge-preserving Lorentz boosts $\eta \rightarrow \eta + a$ are unbroken symmetries;
this condition imposes the following restrictions on (\ref{general}):
\begin{equation}
B_{\omega,\omega'}  = b(\omega) \, \delta_{\omega,\omega'}, \ \ \ \ C_{\omega,\omega'} = 0.
\end{equation}
At this point, we have constructed a family of states parameterized
by an arbitrary function $b(\omega)$; they are associated to the following two-point functions:
\begin{eqnarray}
\W_{b}(x,y) &=&
\frac 1{\pi^2}\int_0^\infty [(b(\omega) +1) e^{-i\omega (\e-\e')} \ +  \cr & &  + \ b(\omega) e^{+i\omega (\e-\e')}]    \,
 K_{i\omega}(m \ex) K_{i\omega}(m e^{\xi'}) {\sinh \pi \omega}\ d\omega  \label{rindler1a}
\end{eqnarray}
The function $b(\omega)$ should be such that  the integral in (\ref{rindler1a}) converges in the sense of distributions.
The choice $b=0$ reproduces the Fulling vacuum \cite{Fulling:1972md,Fulling:1977zs} for the Rindler's Klein-Gordon field.
Taking inspiration from the examples of the previous section, it is now useful to introduce a function $\gamma(\omega)$ such that
\begin{equation}
\b{(\omega)} = {{\frac{e^{-
\frac 12 \gamma(\omega)}}{ 2\sinh(\gamma(\omega)/2)}}};
\end{equation}
so that the two-point function is rewritten as follows:
\begin{equation}
\W_{\gamma}(x,y) =
\frac 1{ \pi^2}\int_0^\infty \left[ \frac{e^{-i\omega (\e-\e') }}{1-e^{-\gamma(\omega)}} +    \frac{e^{i\omega (\e-\e')}}{ e^{\gamma(\omega)}-1}\right]
K_{i\omega}(m \ex) K_{i\omega}(m e^{\xi'}) {\sinh \pi \omega}\ d\omega;  \label{rindler1b}
\end{equation}
in this parametrization the Fulling vacuum corresponds to the choice $\gamma = \infty$.

{It possible to find a choice of $\gamma(\omega)$ such that
the corresponding quantum field theory is fully Poincar\'e
invariant. Hence we study the variation of (\ref{rindler1b})
w.r.t. infinitesimal space translations $e^\xi \, \delta \e = -
\epsilon \, \sinh\e$ and $ e^\xi \delta \xi  = \epsilon \ \cosh
\e$ that map the wedge into itself. Imposing vanishing of the variation and the absence of negative Minkowskian
energies we get the unique solution
$\gamma(\omega) = 2\pi\omega $. In this
case the two-point function $\W_{2\pi}$ can be explicitly identified:
\begin{eqnarray}
{\W}_{2\pi}(x,x')  = \frac
1{\pi^2}\int_0^\infty \ K_{i\omega}(m e^\xi) K_{i\omega}(m
{e^{\xi'}})\cosh{\omega (\pi - i \e-i \e' )} d\omega =  \cr
= \frac 1{2 \pi}  K_{0}\left(m \sqrt{e^{2\xi}+e^{2\xi'}-
2e^{\xi+\xi'}\cosh (\eta-\eta')}\right) = \frac 1{2 \pi}
K_{0}\left(m \sqrt{-(x-x')^2} \right) .
\end{eqnarray}
We have recovered the standard Poincar\'e invariant
quantization of a massive Klein-Gordon field. Unruh's
interpretation follows. The value  of our ``extended canonical quantization'' appears here clearly, as it may work
in situations where the extension to a larger
manifold is not as obvious as in the Rindler case.

A remarkable difference  between the system (\ref{minkmodes})
and the system (\ref{rinmodes}) is that the modes
(\ref{rinmodes}) cannot be distinguished from their complex
conjugates by their behavior at imaginary infinity; {both, indeed, have
the same analyticity properties in the imaginary time
variable simply
because the time coordinate $\eta$ is periodic in the imaginary
part.  This is one sort of circumstance where our generalization
of the canonical scheme proves to be useful. Similar remarks
apply to the previously displayed KMS quantization.
Note also that there is no Poincar\'e invariant}
quantum field theory in the class of models that can be
obtained by standard Bogoliubov transformations (\ref{kgab}) of
the Rindler modes (\ref{rinmodes}) {\em within the Rindler wedge}.
}

There are other theories having a special status in the family (\ref{rindler1a}),
which we recall is already a subset of the general family (\ref{general}).
The most noticeable example is the one-parameter family of states identified by the choice
\begin{equation}
\gamma(\omega) = \beta \omega ,    \ \ \ \ \beta > 0.
\end{equation}
Let us  write the corresponding two-point function explicitly:
\begin{equation}
\W_{\beta}(x,y) =
\frac 1{ \pi^2}\int_0^\infty \left[ \frac{e^{-i\omega (\e-\e') }}{1-e^{-\beta\omega}} +    \frac{e^{i\omega (\e-\e')}}{ e^{\beta \omega}-1}\right]
K_{i\omega}(m \ex) K_{i\omega}(m e^{\xi'}) {\sinh \pi \omega}\ d\omega.  \label{rindler1c}
\end{equation}
Since $K_{i\omega} = K_{-i\omega}$  and since $|K_{i\omega}(\rho) K_{i\omega}(\rho') {\sinh \pi \omega}|$ is bounded at infinity in the $\omega$ variable,
one can immediately  check that $\W_{\beta}(x,y)$ verifies the KMS analyticity and periodicity  properties in imaginary time \cite{haag,Bros:1992ey} at inverse temperature
$\beta$.  These states are precisely  the KMS states in Rindler space, which have been introduced and   characterized in  \cite{kay,kay2}.
The special value $\beta = 2\pi$ has also been identified \cite{kay} with the restriction to the wedge of the Wightman vacuum on the basis
of the Bisognano-Wichmann and Reeh-Schlieder theorems. Our proof follows just by enforcing the requirement that the wedge preserving
translation be an exact symmetry.

\section{Open de Sitter space \label{sect:opendeSitt}}

{This model may be
described by the the metric}
\begin{equation}ds^2 = d{t}^2 -  \sinh^2
{t} \, \frac{(d r^2+ d\x^2_1+ d\x^2_2)}{ r^{2}}, \;t>0,\;r>0,\;
\x_1,\x_2 \in {\Bbb R}. \label{open}\end{equation} {Such a metric defines an instance of a Lema\^{\i}tre-Friedmann
hyperbolic space \cite{MS1} on the de Sitter hyperboloid, which we call here ``open de Sitter space{\rdq}. The coordinate system} used in
(\ref{open}) describes only part of a larger manifold and,
correspondingly, the spacelike surfaces  $t$=const are not
complete Cauchy surfaces for the complete manifold i.e. for the
de Sitter universe itself.

Consider now the massive
Klein-Gordon equation in the open de Sitter universe. A system
of function solving (\ref{kgbase})  is the following:
\begin{equation}
u_{iq,\k}(x)  =  -
\frac{iq}{(2 \pi)^{\frac{3}2}} \sqrt{\frac{\pi}{2\,\sinh \pi q}}
\frac{P^{iq}_{-\frac{1}{2} + i\nu }(\cosh {t} )}{\sinh t} \,\left(\frac{\left({\x}-{\k}  \right)^2}{2 r} + \frac{ r}{2} \right)^{i
{q}- 1}.
\end{equation}
$P$ is the associate Legendre function of the first kind
\cite{BAT2}; the parameter $\nu$ is
related  to the mass by
$m^2 = \frac 94 + \nu^2$. }
We then apply
the most general quantization scheme given in Eq.
(\ref{general})
to find in that class the so-called ``Euclidean''
\cite{gibbons,BM} fully de Sitter invariant, theory.  For
$m^2>2$ the following operators
\[
B_{q\k q'\k'} = \frac{e^{-\pi q}\delta_{qq'}\delta_{\k\k'}}{2\sinh \pi q},
\;\;\;
C_{q\k q'\k'} = \frac{\Gamma\left(\frac{1}{2}-i\nu-iq  \right)
\Gamma\left(\frac{1}{2}+i\nu -i q  \right)\delta_{qq'}\delta_{\k\k'}}{{2\sinh \pi q}\Gamma\left(\frac{1}{2}-i\nu  \right)
\Gamma\left(\frac{1}{2}+i\nu   \right)}
\]
give the {answer we seek}:
\begin{eqnarray}
&& \W(x,x')= \int_0^\infty dq \int d\k \left[\frac{e^{\pi q}u_{iq,\k}(x) u_{iq,\k}^*(x')}{2\sinh \pi q}
+ \frac{e^{-\pi q}u_{iq,\k}^*(x) u_{iq,\k}(x')}{2\sinh \pi
q}\right.+ \cr &&  + 2\Re\int_0^\infty dq  \frac{\Gamma\left(\frac{1}{2}-i\nu-iq  \right)
\Gamma\left(\frac{1}{2}+i\nu -i q  \right)}{{2\sinh \pi q}\Gamma\left(\frac{1}{2}-i\nu  \right)
\Gamma\left(\frac{1}{2}+i\nu   \right)}\int d\k u_{iq,\k}(x) u_{iq,\k}(x')=
\cr &&
= \frac{\Gamma\left(\frac{3}{2} +i\nu\right)
            \Gamma\left(\frac{3}{2} -i \nu\right)}
{8\pi^{2}} (\zeta^2-1)^{-\frac{1}{2}}\, P^{-1}_{-\frac{1}{2} +
i\nu}(\zeta). \label{Pdef}
\end{eqnarray}
where $ \zeta =
 ({2 r r'})^{-1}{\sinh t \sinh t'[\left({\x}-{\x'}  \right)^2+ r^2
+{ r'}^2]} -\cosh t \cosh t'$. For $m^2<2$ the Euclidean vacuum
can be recovered by adding a   contribution $S(x,x')$ which is
classical and hence need not be square integrable as discussed
after our general formula for quantization Eq. (\ref{general});
here is an example where the $L^2$ canonical quantization
fails. The implications of this novelty in quantum field theory
and statistical physics { deserve a full specific discussion
and} will be explored elsewhere \cite{usexp}.

We see once again that our quantization scheme allows to work
solely within the known coordinate patch; no extension of the
``physical'' space to ``elsewhere'' has been used. Performing such
an extension \cite{sasaki,MS} indeed yields the same result,
but with much more effort. Furthermore, in cases where the
geometry is not so well understood as in the de Sitter case,
the way to describe the extension (if any)  may be not under
control: the power of our  generalization of canonical
quantization in curved space-times  appears here clearly.

\section{Summary and concluding remarks\label{sec:concl}}

There is much more flexibility in canonical quantization than
is usually believed. The  simple but very general modification of the
standard formalism  that we have described in Sect. \ref{sect:phidetout}
opens a vast class of new possibilities for constructing
canonical fields by means of the Fock construction.
The modification amounts to considering the most general quadratic combination
of a given set of modes that is compatible with a given
commutator function, or, equivalently, solves the fundamental split Equation (\ref{CR}).

Our scheme produces for instance an original and simple
construction of the thermal equilibrium states, as long as a
wealth of other similar unexplored possibilities:
pure states and mixed as well are encompassed in our construction.

Another example which has even more interesting features is
quantum field theory in the Rindler wedge. Here, the important
characteristic of our approach is its ability to reconstruct
the standard Poincar\'e invariant vacuum working solely within
the Rindler wedge. Only local invariance is required to  get the globally invariant vacuum, with no need to consider analytic continuation to (or from) whatever ``external'' complementary space which may be introduced.

We have briefly discussed the de Sitter case in our last
example. Also in this case the coordinate system gives access only to a
portion of the manifold; while canonical quantization in these
coordinates (plus Bogoliubov transformations)
produces theories that are not de Sitter invariant, application
of our procedure gives rise to the preferred de Sitter
invariant theory \cite{gibbons,BM}.
When the mass is lower than a critical value there are also
non-standard non-square integrable modes which come into the
play. These modes were known to exist, but their physical
relevance was quite uncertain. We have seen here that such contributions
are not at variance with the principles of quantum mechanics
since they do not contribute to the commutator.
This comforts their relevance, and is another novel feature which will be
discussed in detail in a further work  \cite{usexp}.

In all these cases,  our procedure allows the
construction of the same fundamental state in various
coordinate systems also where the usual canonical quantization is not
able of doing so. We expect such kind of invariance to exist
on  general grounds, reflecting the requirement that the
description of a given physical state be possible
in local system of coordinates. To achieve this goal our generalization of the standard
canonical quantization procedure seems relevant and opens the way for a
general study of these invariance properties.
We have insisted in using the words ``canonical quantization'' because we are always looking for Fock representations
of the canonical commutation relations. Our formalism goes however beyond the standard formalism in that it produces pure and mixed states on an equal footing; also it allows negative energy states always within the limits of canonicity; non-$L^2$ contributions are  also  allowed (these contributions are of classical nature).
That is why it might be called  {\it ``extended canonical quantum field theory''}.
We leave to further study the question whether this
extension encompasses all the possible
Fock representations of the Klein-Gordon field algebra.

\end{document}